\documentclass[preprint,superscriptaddress,showpacs,tightenlines]{revtex4}
\usepackage{amssymb}
\usepackage{amsmath}
\usepackage{graphicx,bm}

\input{tcilatex}

\begin{document}

\title{Quantum interference effects in a system of two tunnel point-contacts
in the presence of single scatterer: simulation of a double-tip STM
experiment}
\author{N.V. Khotkevych}
\affiliation{B.I. Verkin Institute for Low Temperature Physics and Engineering, National
Academy of Sciences of Ukraine, 47, Lenin Ave., 61103, Kharkov, Ukraine.}
\author{Yu.A. Kolesnichenko}
\affiliation{B.I. Verkin Institute for Low Temperature Physics and Engineering, National
Academy of Sciences of Ukraine, 47, Lenin Ave., 61103, Kharkov, Ukraine.}
\author{J.M. van Ruitenbeek}
\affiliation{Kamerlingh Onnes Laboratorium, Universiteit Leiden, Postbus 9504, 2300
Leiden, The Netherlands.}

\begin{abstract}
The conductance of systems containing two tunnel point-contacts and a single
subsurface scatterer is investigated theoretically. The problem is solved in
the approximation of $s$-wave scattering giving analytical expressions for
the wave functions and for the conductance of the system. Conductance
oscillations resulting from the interference of electron waves passing
through different contacts and their interference with the waves scattered
by the defect are analyzed. The prospect for determining the depth of the
impurity below the metal surface by using the dependence of the conductance
as a function of the distance between the contacts is discussed. It is shown
that the application of an external magnetic field results in Aharonov-Bohm
type oscillations in the conductance, the period of which allows detection
of the depth of the defect in a double tip STM experiment.
\end{abstract}

\maketitle

With the further development of scanning tunnelling microscopy (STM) it has
become clear that a single STM-probe is often not enough for obtaining
information on the detailed characteristics of the surface under
investigation. A logical development of the one-tip approach is a dual-tip
experimental setup, which can provide us with richer information than
conventional single-probe STM. Despite the apparent technical complexity of
the dual-tip STM (DSTM) in comparison with standard STM several groups have
demonstrated successful solutions for such refinement of the STM-technology 
\cite{Jasninsky2006,Okamoto,YiKaya,Grube}.

DSTM can be realized in different ways. For example, it can be a spatially
extended STM tip with two protrusions, each ending in a cluster or a single
atom \cite{Flatte}. A second approach is a coaxial beetle-type double-tip
STM design that looks advantageous in retaining the standard STM stability 
\cite{Jasninsky2008}. The most versatile DSTM comprises two individual tips,
which can be driven independently. In this case the distance between the
tips is limited in principle only by a parameter such as the characteristic
tip radius \cite{Okamoto}. Another original example of the DSTM was proposed
in \cite{Byers}, where one contact can be created directly on the surface,
while the other one was the STM-tip itself.

For DSTM experiments with two independent probes there are different
possibilities for applying voltages to the tunnelling contacts. There are
two basic circuit designs: In the first one electrons are emitted from the
first contact and then gathered at the second, i.e. the current flows from
one contact to the other through the surface being probed \cite{Niu,Dana}.
This method allows capturing a trans-conductance map, and in addition allows
the implementation of three-terminal ballistic electron emission
spectroscopy (BEES) without introduction of macroscopically bounded contacts 
\cite{YiKaya}. In the second basic scheme proposed in Ref.~\cite{Flatte} the
bias is applied between the two tips and the sample, i.e. the current flows
from two contacts into the sample.

Subsurface defects, adatoms, and steps on the metal surface result in the
appearance of Friedel-like oscillations in the STM conductance $G=dI/dV$ - a
nonmonotonic dependence of $G$ with the distance between the STM tip and the
defect $r_{0}$ (for a review see \cite{AKR}). The study of this dependence
can be used for the detection of buried defects and for investigation of
their characteristics. Methods for determining defect positions below a
metal surface using a single tip STM have been proposed before: this can be
achieved using the period of oscillation of the conductance as a function of
bias \cite{Kobayashi,Avotina2005} or by exploiting the interference pattern
of conductance as a function of position, $G\left( r_{0}\right) $, which is
very pronounced for open directions of Fermi surface \cite%
{Avotina06,Avotina08nm,Wiesmann}. These approaches are very suitable for the
surfaces of simple metals, such as the noble metals, but application to
conductors having a more complicated Fermi surface geometries will be
difficult and has not yet been explored.

In the present work we examine the case of injection of electrons to the
surface by the first and the second contacts simultaneously. We consider
this realization of a double-tip experiment as a natural refinement of the
single-tip STM problem for the study of single defects buried under the
metal surface \cite{Kobayashi,Avotina2005,Wiesmann}.

\begin{figure}[tbp]
\includegraphics[width=10cm,angle=0]{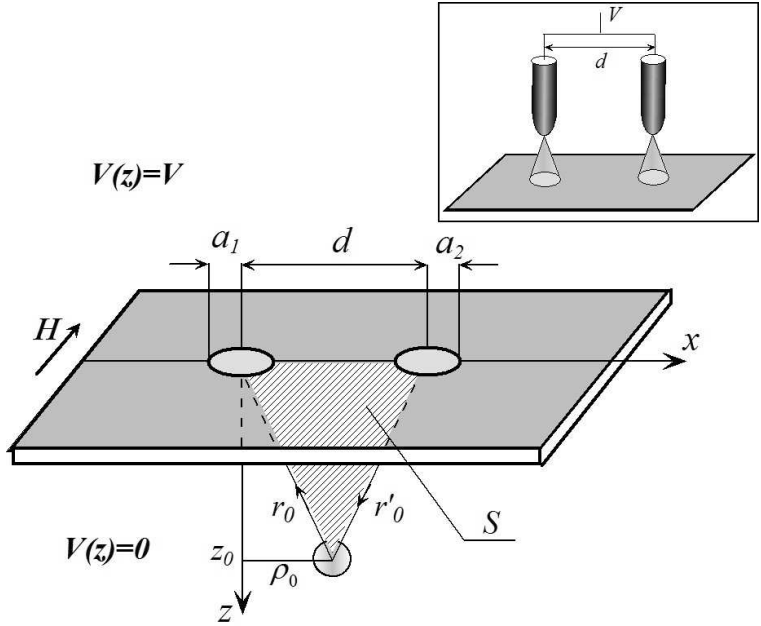}
\caption{Schematic arrangement of the system of two tunnel contacts,
modelled as two orifices in an infinitely thin interface between conducting
half-spaces. The inset shows the equivalent circuit with two STM tips, which
provide the electron tunnelling paths through small areas with
characteristic radii $a_{1}$ and $a_{2}$. }
\label{model}
\end{figure}

The idea of using multiple tunnelling contacts for determining the depth and
location of impurities under a metal or semiconductor surface has been
expressed earlier in Ref.~\cite{Niu}. The paper by Niu \textit{et al.}, Ref.~%
\cite{Niu}, proposes a method for determining the desired depth by measuring
the trans-conductance between two tips of the dual-tip scanning tunneling
microscope. In the present paper we propose a different approach, namely, by
measuring the phase change $\Delta \vartheta $\ in the conductance
oscillations as a function of the distance between two STM tips $d$. Such
phase changes can be measured experimentally with great precision. We show
that $\Delta \vartheta $ can be expressed in terms of the distance $d$ (in
units of the Fermi wave vector $k_{\mathrm{F}}$), the position of the defect
in the plane parallel to surface plane $\rho _{0}$, which is easily defined
experimentally, and the unknown depth of the defect $z_{0}$. Thus by
measurement of $\Delta \vartheta \left( d\right) $ it is possible to
determine the depth of the buried impurity. The procedure of defining $z_{0}$
is further simplified when a magnetic field $\mathbf{H}$ is applied to the
system. In this case the STM conductance $G$ undergoes Aharonov-Bohm type
oscillations. These oscillations result from the quantization of the
magnetic flux through the area formed by the electron trajectories from the
contacts to the defect and the line connecting the contacts (Fig.\ref{model}%
). For a weak magnetic field the electron trajectories and the line
connecting the contacts form a triangle, and from its area $S$ the defect
depth can be found easily.

As a model for the double-tip STM geometry we consider two metal half-spaces
separated by an infinitely thin nonconducting interface at $z=0$, which
contains two small regions (contacts) that allow electron tunnelling (see
Fig.\ref{model}). The origin of the coordinate system $\mathbf{r}=0$ is
chosen in the center of the first contact. The $x$ axis is directed along
the the line connecting the contacts. For the potential barrier in the plane 
$z=0$ we use the function \cite{Kulik} 
\begin{equation}
U\left( \mathbf{r}\right) =U_{0}f\left( \mathbf{\rho }\right) \delta \left(
z\right) .  \label{U}
\end{equation}%
In our case $f\left( \mathbf{\rho }\right) $ describes two "windows" for
electron tunneling and the reciprocal function $f^{-1}\left( \mathbf{\rho }%
\right) $ can be presented as a sum of two terms 
\begin{equation}
f^{-1}\left( \mathbf{\rho }\right) =\chi \left( \rho /a_{1}\right) +\chi
\left( \left\vert \mathbf{d}-\mathbf{\rho }\right\vert /a_{2}\right) ,\quad
2\pi \int\limits_{0}^{\infty }dxx\chi \left( x\right) =1,  \label{1/f}
\end{equation}%
where $\chi \left( x\right) \simeq 1$ for $x\lesssim 1,$ and $\chi \left(
x\right) \ll 1$ for $x\gg 1,$ $a_{1,2}$ are the characteristic radii of the
contacts, $\mathbf{\rho }$ is the component of the vector $\mathbf{r}$
parallel to the plane $z=0,$ $\mathbf{d}$ is a two dimensional radius vector
from the center of first contact to the center of second one. The absolute
value $d$ is the distance between contacts, assuming that this is smaller
than the shortest relaxation length.

In the vicinity of the contacts a single defect is placed described a short
range potential $D(\mathbf{r})$, 
\begin{equation}
D\left( \mathbf{r}\right) =gD_{0}\left( \left\vert \mathbf{r-r}%
_{0}\right\vert \right) ,  \label{D(r)}
\end{equation}%
where $g$ is the constant of interaction of the electrons with the defect,
and $D_{0}(\left\vert \mathbf{r-r}_{0}\right\vert )$ is a spherically
symmetric function localized within a region of characteristic radius $r_{D}$
centered at the point $\mathbf{r}=\mathbf{r}_{0}$, which satisfies the
normalization condition 
\begin{equation}
4\pi \int\limits_{0}^{\infty }dr^{\prime }r^{\prime 2}D_{0}(r^{\prime })=1.
\label{intD=1}
\end{equation}

For calculation of the conductance $G$ we proceed as before. The probability
density current is found by using the wave function $\psi \left( \mathbf{r}
\right) $ for the electrons tunnelling through the potential barrier in the
plane of the orifices. The total electric current $I$ in the system is
calculated by integrations over electron momenta and over a real-space
surface overlapping the contacts. We will take the temperature to be zero,
and assume a small applied voltage $V$ such that we stay in the linear
regime of Ohm's law, $I=GV$. Under these assumptions the conductance $G$ can
be written as%
\begin{equation}
G=\frac{e^{2}\hbar }{m^{\ast }}\nu \left( \varepsilon _{\mathrm{F}}\right)
\int\limits_{S_{\mathrm{F}},v_{z}>0}d\Omega _{\mathbf{p}}\int\limits_{S}d%
\Omega r^{2}\func{Im}\left[ \psi ^{\ast }\left( \mathbf{r}\right) \nabla
\psi \left( \mathbf{r}\right) \right] .  \label{G}
\end{equation}%
In Eq.(\ref{G}) $m^{\ast }$ is the effective electron mass, $\nu \left(
\varepsilon _{\mathrm{F}}\right) $ is the electron density of states at the
Fermi level, $d\Omega $ and $d\Omega _{\mathbf{p}}$ are solid angles in the
real and momentum spaces, respectively. As the surface for space integration
we choose a half-sphere of radius $r$, larger than distance between the
contacts $d$ and centered at the center of first contact, $r=0,$ and
covering the contacts in the lower half-space, $z>0.$ The integration over
the directions of the momentum over the Fermi surface $S_{\mathrm{F}}$ is
carried out for electrons tunnelling and having a positive projection $v_{z}$
of the electron velocity on the contact axis $z.$ As a consequence of the
conservation of total current the integral over $d\Omega $ does not depend
on the length we choose for the radius $r.$

The electron wave function $\psi \left( \mathbf{r}\right) $ satisfies the
Schr\"{o}dinger equation%
\begin{equation}
\left[ \nabla ^{2}+\frac{2m^{\ast }}{\hbar ^{2}}\left( \varepsilon -D(%
\mathbf{r})\right) \right] \psi \left( \mathbf{r}\right) =0,  \label{Schrod}
\end{equation}%
subject to the boundary conditions of continuity and of the jump of its
derivative at $z=0$. In Ref.\cite{Kulik} a solution of Eq.(\ref{Schrod}) was
found for an arbitrary function $f\left( \mathbf{\rho }\right) $, in the
limit of weak tunnelling, $1/U_{0}\rightarrow 0$, and for a purely ballistic
contact (no defects present), 
\begin{equation}
\psi _{0}\left( \mathbf{\rho },z\right) =\frac{-ik_{z}\hbar ^{2}}{\left(
2\pi \right) ^{2}m^{\ast }U_{0}}\int\limits_{-\infty }^{\infty }d\mathbf{%
\varkappa }^{\prime }e^{i\mathbf{\varkappa }^{\prime }\mathbf{\rho }%
}e^{ik_{z}^{\prime }z}\int\limits_{-\infty }^{\infty }d\mathbf{\rho }%
^{\prime }\frac{e^{i\left( \mathbf{\varkappa }-\mathbf{\varkappa }^{\prime
}\right) \mathbf{\rho }^{\prime }}}{f\left( \mathbf{\rho }^{\prime }\right) }%
,  \label{wfKulik}
\end{equation}%
where $k_{z}^{^{\prime }}=\sqrt{\varkappa ^{2}+k_{z}^{2}-\varkappa ^{\prime
2}}$, and $\mathbf{\varkappa }$ and $k_{z}$ are the components of the vector 
$\mathbf{k}$ parallel \ and perpendicular to the interface, respectively. As
a special case the authors of Ref.\cite{Kulik} considered a system of
several orifices with different radii.

The characteristic radius of the region through which the electrons tunnel
from the STM tip into the sample has sub-atomic size $(a\simeq 0.1\mathring{A%
})$ while the Fermi wave vector is $k_{\mathrm{F}}\simeq 1$\AA $^{-1}.$ By
using the condition $k_{\mathrm{F}}${}$a_{1,2}\ll 1$ we find, after
integrating over $\mathbf{\varkappa }^{\prime }$ in Eq.(\ref{wfKulik}),

\begin{equation}
\psi _{0}\left( \mathbf{r}\right) =\frac{it\left( k_{z}\right) }{2}\left[
\left( ka_{1}\right) ^{2}\frac{z}{r}h_{1}^{\left( 1\right) }(kr)+\left(
ka_{2}\right) ^{2}\frac{z}{r^{\prime }}h_{1}^{\left( 1\right) }(kr^{\prime
})e^{i\mathbf{\varkappa d}}\right]  \label{wf_asym}
\end{equation}%
where $h_{1}^{\left( 1\right) }(kr)$ is the spherical Bessel function of the
first order, $r^{\prime }=\left\vert \mathbf{r-d}\right\vert ,$ and 
\begin{equation}
t\left( k_{z}\right) =\frac{\hbar ^{2}k_{z}}{im^{\ast }U_{0}},  \label{t}
\end{equation}%
is the transmission amplitude of the electron wave function passing through
a homogeneous barrier. Note that in the limit $k_{\mathrm{F}}${}$%
a_{1,2}\rightarrow 0$ the result of Eq.~(\ref{wf_asym}) does not depend on
the concrete form of the function $\chi \left( \rho /a\right) $ in Eq.~(\ref%
{1/f}) and the wave function Eq.~(\ref{wf_asym}) as well as the conductance
of the system are expressed in terms of the effective areas of the contacts, 
$\pi a_{1,2}^{2}.$ 
\begin{figure}[tbp]
\includegraphics[width=10cm,angle=0]{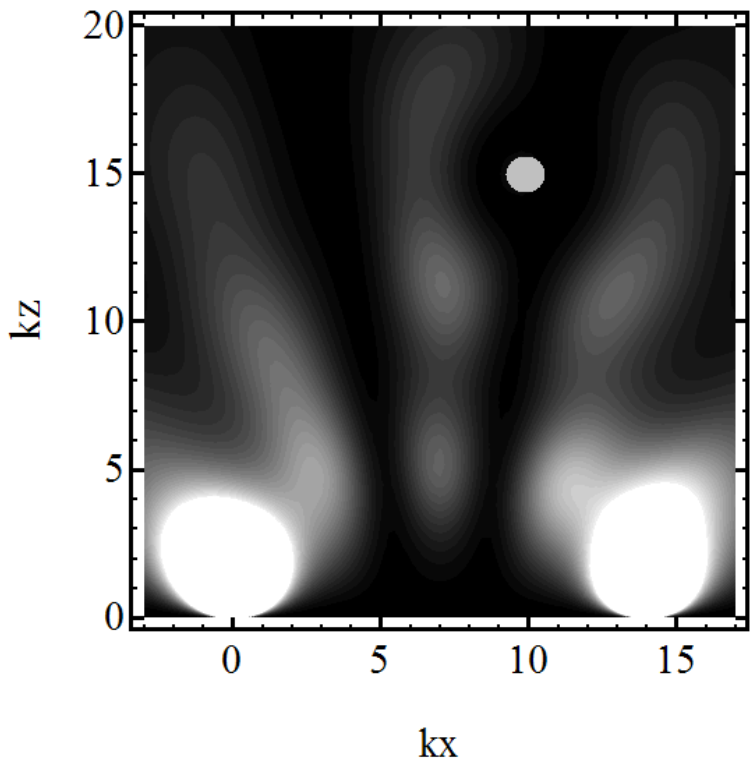}
\caption{Squared modulus of the wave function (\protect\ref{psi}). The
defect sits at $kr_{0}=(10,0,15),$ the distance between the contacts $kd=14,$
the scattering phase shift is $\protect\delta _{0}=1.5$. }
\label{space_distr}
\end{figure}

The effect of electron scattering by the short-range potential can be taken
into account by the method proposed in Ref.\cite{Grenot}. If the radius of
action $r_{D}$ of the potential $D(\mathbf{r})$ is of the order of Fermi
wave length $\lambda _{\mathrm{F}},$ in the region of the defect $\left\vert 
\mathbf{r}- \mathbf{r}_{0}\right\vert \lesssim r_{D}$ the wave function $%
\psi \left( \mathbf{r}\right) $ can be taken as a constant $\psi \left( 
\mathbf{r} _{0}\right) ,$ as for a $\delta $-function. Under this
approximation Eq.~(\ref{Schrod}) takes the form of a non-homogeneous
equation with the right-hand member being $\frac{2m^{\ast }}{\hbar ^{2}}%
D\psi \left( \mathbf{r}_{0}\right) .$ In the limit $1/U_{0}\rightarrow 0$ a
solution of this equation can be expressed in terms of the solution of the
homogeneous equation (see Eqs.~(\ref{wfKulik}), (\ref{wf_asym})) and the
retarded electron Green's function of Eq.(\ref{Schrod}) for the
semi-infinite half-space

\begin{equation}
\psi \left( \mathbf{r}\right) =\psi _{0}\left( \mathbf{r}\right) +\frac{%
2m^{\ast }}{\hbar ^{2}}T\left( k\right) \left[ G_{0}^{\left( +\right)
}\left( \left\vert \mathbf{r}-\mathbf{r}_{0}\right\vert \right)
-G_{0}^{\left( +\right) }\left( \left\vert \mathbf{r}-\widetilde{\mathbf{r}}%
_{0}\right\vert \right) \right] \psi _{0}\left( \mathbf{r}_{0}\right) ,
\label{psi}
\end{equation}%
where $\widetilde{\mathbf{r}}_{0}=\left( \mathbf{\rho }_{0},-z_{0}\right)$. $%
T\left( k\right) $ is the scattering matrix, which for a short-range
scatterer can be expressed in terms of the s-wave scattering phase shift $%
\delta _{0}$ \cite{Avotina2008}

\begin{equation}
T\left( k\right) =\frac{-\pi \hbar ^{2}\left( e^{2i\delta _{0}}-1\right) }{%
m^{\ast }ik\left( 1+\frac{1}{4ikz_{0}}\left( e^{2i\delta _{0}}-1\right)
e^{2ikz_{0}}\right) }.  \label{T}
\end{equation}
The Green's function 
\begin{equation}
G_{0}^{\left( +\right) }\left( x\right) =-\frac{\exp \left( ikx\right) }{%
4\pi x}
\end{equation}%
is the retarded Green's function of a free electron. The phase shift $\delta
_{0}$ is determined by the scattering strength $g$ as, 
\begin{equation}
e^{i\delta _{0}}\sin \delta _{0}=-\frac{m^{\ast }kg}{2\pi \hbar ^{2}}\left(
1-\frac{8\pi m^{\ast }g}{\hbar ^{2}}\int\limits_{0}^{\infty }drG_{0}^{\left(
+\right) }\left( r\right) D\left( r\right) \right) ^{-1}.
\end{equation}

Fig.\ref{space_distr} illustrates the spacial variation of the wave function
(\ref{psi}) for the case when the contacts and the scatterer are all placed
in the plane $y=0$. The interference of election waves passing through
different contacts and their interference with the waves scattered by the
defect are clearly visible. In order to make the effects more visible we
used in Fig.\ref{space_distr} a large value for the scattering phase $\delta
_{0}=1.5,$ which is acceptable only for Kondo resonance scattering by a
magnetic impurity (see, for example, \cite{Abrikosov}). The grey circle
round the point $\mathbf{r}=\mathbf{r}_{0}$ in Fig.\ref{space_distr} is the
region in which the Eq.(\ref{psi}) is not valid because of divergence of the
Green function.

Substituting the wave function (\ref{psi}) into the general expression for
the conductance $G$ (\ref{G}) we find%
\begin{equation}
G=G_{c}\left( a_{1},a_{2}\right) +G_{osc},  \label{FullCond}
\end{equation}%
where $G_{c}$ is the conductance of the double contact system in the absence
of the defect%
\begin{equation}
G_{c}\left( a_{1},a_{2}\right) =G_{0}\left( a_{1}\right) +G_{0}\left(
a_{2}\right) +G_{12},  \label{G_c}
\end{equation}%
$G_{0}$ is an inherent conductance of the single contact \cite{Kulik} 
\begin{equation}
G_{0}\left( a\right) =\left\vert t\left( k_{\mathrm{F}}\right) \right\vert
^{2}\frac{e^{2}\left( k_{\mathrm{F}}a\right) ^{4}}{36\pi \hbar },  \label{G0}
\end{equation}%
and $G_{12}$ takes into account the interference of electron waves passing
through different contacts 
\begin{equation}
G_{12}=\left\vert t\left( k_{\mathrm{F}}\right) \right\vert ^{2}\frac{%
e^{2}\left( k_{\mathrm{F}}a_{1}\right) ^{2}\left( k_{\mathrm{F}}a_{2}\right)
^{2}}{18\pi \hbar }f^{2}\left( k_{\mathrm{F}}d\right) .
\end{equation}%
Here we introduced the notation%
\begin{equation}
f\left( x\right) =\frac{3j_{1}\left( x\right) }{x},
\end{equation}%
$j_{1}\left( x\right) $ the spherical Bessel function of the first kind such
that $f\left( 0\right) =1.$ The second term in Eq.(\ref{FullCond}), $%
G_{osc}, $ describes the quantum interference resulting from the scattering
of the electrons by the defect 
\begin{equation}
G_{osc}\left( r_{0},d\right) =G_{0}\left( a_{1}\right) \Gamma \left( \mathbf{%
r}_{0}\right) +G_{0}\left( a_{2}\right) \Gamma \left( \mathbf{r}_{0}^{\prime
}\right) +2G_{12}\Psi \left( \mathbf{r}_{0},\mathbf{r}_{0}^{\prime }\right)
/f\left( k_{\mathrm{F}}d\right) ,  \label{Gosc}
\end{equation}%
where $\mathbf{r}_{0}^{\prime }=\left( \mathbf{\rho }_{0}-\mathbf{d}%
,z_{0}\right) $, and $r_{0}^{\prime }$ is the distance between the defect
and the second contact. The functions $\Gamma \left( \mathbf{r}_{0}\right) $
and $\Gamma \left( \mathbf{r}_{0}^{\prime }\right) $ take into account the
effect of interference of electron waves passing through the contact and
returning to the same contact after scattering by the defect, 
\begin{equation}
\Gamma \left( \mathbf{r}\right) =\frac{1}{F\left( z\right) }\sin \delta _{0}%
\frac{z^{2}}{r^{2}}\left[ 12j_{1}\left( k_{\mathrm{F}}r\right) \gamma \left(
k_{\mathrm{F}}r\right) +6\left( 1-j_{0}\left( 2k_{\mathrm{F}}z\right)
\right) \left( k_{\mathrm{F}}r\right) ^{-4}\left( \left( kr\right)
^{2}+1\right) \sin \delta _{0}\right] ,  \label{gamma}
\end{equation}%
where 
\begin{equation}
F\left( z\right) =1+2\sin \delta _{0}\left[ \left( \frac{1}{2\left( 2k_{%
\mathrm{F}}z\right) ^{2}}-j_{0}\left( 2k_{\mathrm{F}}z\right) \right) \sin
\delta _{0}-y_{0}\left( 2k_{\mathrm{F}}z\right) \cos \delta _{0}\right] ,
\end{equation}%
and 
\begin{equation}
\gamma \left( \mathbf{r}\right) =-y_{1}\left( k_{\mathrm{F}}r\right) \cos
\delta _{0}+\sin \delta _{0}\left\{ j_{1}\left( k_{\mathrm{F}}r\right)
\left( j_{0}\left( 2k_{\mathrm{F}}z\right) -1\right) +y_{0}\left( 2k_{%
\mathrm{F}}z\right) y_{1}\left( k_{\mathrm{F}}r\right) \right\} .
\label{Gamma}
\end{equation}%
In the last term in Eq.(\ref{Gosc}) $\Psi \left( \mathbf{r}_{0},\mathbf{r}%
_{0}^{\prime }\right) $ describes the interference of electron waves that
arrive at the other contact after scattering by the defect, 
\begin{eqnarray}
\Psi \left( \mathbf{r},\mathbf{r}^{\prime }\right) &=&F^{-1}\sin \delta _{0}%
\frac{z^{2}}{rr^{\prime }}\left[ j_{1}\left( k_{\mathrm{F}}r^{\prime
}\right) \gamma \left( \mathbf{r}\right) +j_{1}\left( k_{\mathrm{F}}r\right)
\gamma \left( \mathbf{r}^{\prime }\right) \right. +  \label{Fr1r2} \\
&&+\sin \delta _{0}\left( j_{0}\left( k_{\mathrm{F}}d\right) -j_{0}\left( k_{%
\mathrm{F}}\sqrt{4z^{2}+d^{2}}\right) \right) \times  \notag \\
&&\left. \left( j_{1}\left( k_{\mathrm{F}}r\right) j_{1}\left( k_{\mathrm{F}%
}r^{\prime }\right) -y_{1}\left( k_{\mathrm{F}}r\right) y_{1}\left( k_{%
\mathrm{F}}r^{\prime }\right) \right) \right] ,  \notag
\end{eqnarray}

\begin{figure}[tbp]
\includegraphics[width=10cm,angle=0]{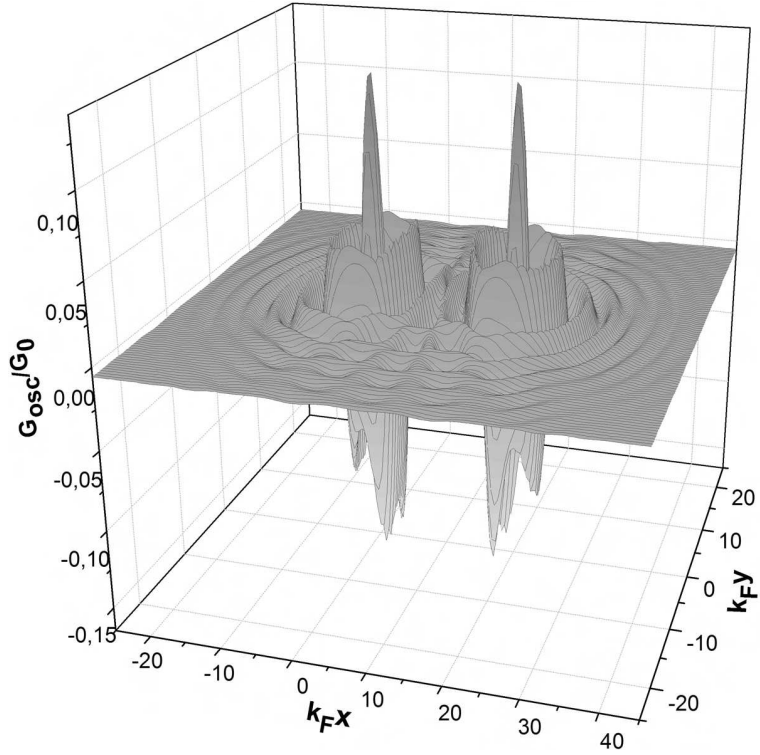}
\caption{Dependence of the normalized oscillatory part of the conductance $%
G_{osc}/G_{0}$ as a function of the defect position $\protect\rho _{0}$ in
the plane parallel to interface $z=0$, $k_{\mathrm{F}}z_{0}=5$ . The
distance between the contacts is taken as $k_{\mathrm{F}}d=20$, and the
scattering phase shift is $\protect\delta _{0}=1.5$. }
\label{osc_pattern}
\end{figure}
For $a_{2}=0$ (i.e., when we have just a single contact) Eq.(\ref{FullCond})
coincides with the expression for the conductance of a tunnel point contact
obtained in Ref.\cite{Avotina2008}. Fig.~\ref{osc_pattern} illustrates the
dependence of the oscillatory part of the conductance (\ref{Gosc}) on the
position of the defect in the plane $z=z_{0}.$ The oscillatory pattern
presented in Fig.~\ref{osc_pattern} represents an image which could be
obtained by DSTM when mapping the tunnelling conductance in the vicinity of
the subsurface defect.

The general formula for the conductance (\ref{FullCond}) can be simplified
for large distances between the contacts and the defect, $%
r_{0},r_{0}^{\prime }\gg 1/k_{\mathrm{F}}$, and for a weak scattering
potential $\delta _{0}\simeq -gm^{\ast }k_{\mathrm{F}}/2\pi \hbar ^{2}\ll 1$%
. Under these assumptions the normalized oscillatory part of the
conductance, in the linear approximation in $g$, can be written as 
\begin{equation}
\frac{G_{osc}}{G_{0}}=-6\delta _{0}\frac{z_{0}^{2}}{k_{\mathrm{F}}^{2}}%
\left\{ \frac{1}{r_{0}^{4}}\sin 2k_{\mathrm{F}}r_{0}+\frac{1}{r_{0}^{\prime
4}}\sin 2k_{\mathrm{F}}r_{0}^{\prime }-\frac{2f\left( k_{\mathrm{F}}d\right) 
}{\left( r_{0}r_{0}^{\prime }\right) ^{2}}\sin k_{\mathrm{F}}\left(
r_{0}+r_{0}^{\prime }\right) \right\} ,  \label{OscCond}
\end{equation}%
For simplicity we take here $a_{1}=a_{2}=a$. Equation~(\ref{OscCond}) shows
that in contrast to one tunnel point contact, for which $G_{osc}\sim \sin
2k_{\mathrm{F}}r_{0}$ when $k_{\mathrm{F}}r_{0}\gg 1,$ the oscillatory
dependence of the double contact has a phase shift $\vartheta $ that depends
on the distance between the contacts 
\begin{equation}
\frac{G_{osc}}{G_{0}}\sim \sin \left( 2k_{\mathrm{F}}r_{0}+\vartheta \right)
,
\end{equation}%
where 
\begin{equation}
\vartheta (\rho _{0},z_{0},d)=-\arcsin \left[ \frac{\sin 2\varphi +2f(k_{%
\mathrm{F}}d)\sin \varphi }{\sqrt{2+4f^{2}(k_{\mathrm{F}}d)+4f(k_{\mathrm{F}%
}d)(\cos 2\varphi +2\cos \varphi )}}\right] ,\quad d\ll r_{0},  \label{teta}
\end{equation}%
\begin{figure}[tbp]
\includegraphics[width=10cm,angle=0]{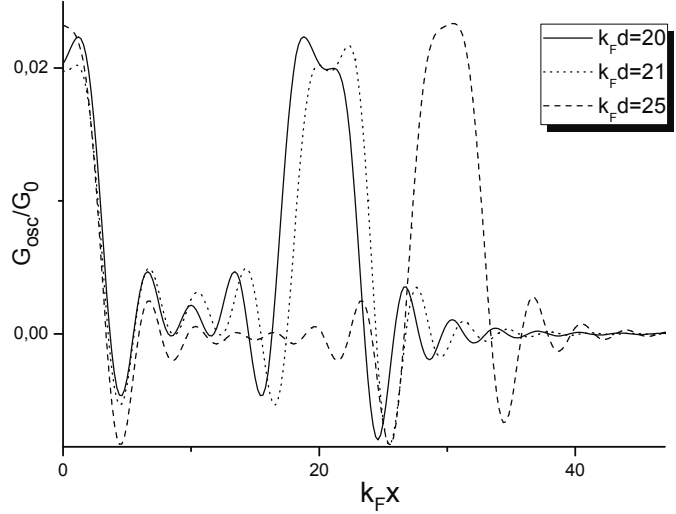}
\caption{Dependencies of the oscillatory part of the conductance on the
coordinate $x_{0}$ of the defect for different distances between the
contacts, for $y_{0}=0$, $k_{\mathrm{F}}z_{0}=5,$ and $\protect\delta %
_{0}=0.1.$}
\label{different_d}
\end{figure}
and $\varphi =\frac{k_{\mathrm{F}}\rho _{0}d}{r_{0}}.$ The defect position
in the plane parallel to the surface, $\rho _{0}$, is known from the
interference pattern of the conductance oscillations (see Fig.\ref%
{osc_pattern}). In principle, the depth of the defect $z_{0}$ may be found
from the experimental data in the following way: changing the distance
between the contacts over a small range $\Delta d\ll d$ leads to the
appearance of an additional phase shift $\Delta \vartheta \left( \rho
_{0},z_{0},d\right) $, which can be defined from the dependence $%
G_{osc}\left( \rho _{0},z_{0},d\right) $, see Fig.~(\ref{different_d}). The
depth $z_{0}$ can be obtained as a numerical solution of the equation

\begin{equation}
\Delta \vartheta =\vartheta _{d}^{\prime }\left( \rho _{0},z_{0},d\right)
\Delta d.  \label{deltateta}
\end{equation}%
\qquad \qquad

Let us now consider applying a magnetic field $\mathbf{H}$ parallel to the
surface of the sample (see Fig.~\ref{model}). If the external magnetic field
is sufficiently weak, such that the radius of the electron trajectories $%
r_{H}=\hbar ck_{\mathrm{F}}/eH\ $ is much larger than the distances between
the the contacts and the impurity, $r_{0}$, $r_{0}^{\prime }$, the magnetic
distortions of the trajectories \cite{Avotina2007} are negligible, i.e. the
trajectories can be considered as straight lines.

Under this condition of $r_{H}\gg r_{0}$, $r_{0}^{\prime }$, the zero-field
wave-function $\psi \left( \mathbf{r}\right) $ acquires an additional phase:

\begin{equation}
\tilde{\psi}\left( \mathbf{r}\right) =\psi \left( \mathbf{r}\right) \exp
\left( \frac{ie}{c\hbar }\int\limits_{0}^{\mathbf{r}}\mathbf{A}\left( 
\mathbf{r}^{\prime \prime }\right) d\mathbf{r}^{\prime \prime }\right) ,
\label{psi_magn}
\end{equation}%
and the Green function similarly takes the form \cite{Aharonov-Bohm}:

\begin{equation}
\tilde{G}\left( \mathbf{r,r}_{0}\right) =G\left( \mathbf{r,r}_{0}\right)
\exp \left( \frac{ie}{c\hbar }\int\limits_{\mathbf{r}_{0}}^{\mathbf{r}}%
\mathbf{A}\left( \mathbf{r}^{\prime \prime }\right) d\mathbf{r}^{\prime
\prime }\right) .
\end{equation}%
Here, $\mathbf{A}\left( \mathbf{r}\right) $ is the vector potential of the
magnetic field.

On account of this change in the wave function (\ref{psi_magn}) the formula
for the conductance $G$ is modified and takes the form:

\begin{equation}
G=G_{c}\left( a_{1},a_{2}\right) +\tilde{G}_{osc}\left( r_{0},d,H\right) ,
\label{FullCond_magn}
\end{equation}

\begin{equation}
\tilde{G}_{osc}\left( r_{0},d,H\right) =G_{0}\left( a_{1}\right) \Gamma
\left( \mathbf{r}_{0}\right) +G_{0}\left( a_{2}\right) \Gamma \left( \mathbf{%
r}_{0}^{\prime }\right) +2G_{12}\tilde{\Psi}\left( \mathbf{r}_{0},\mathbf{r}%
_{0}^{\prime }\right) /f\left( k_{\mathrm{F}}d\right) ,  \label{Gosc_magn}
\end{equation}

\begin{eqnarray}
\tilde{\Psi}\left( \mathbf{r},\mathbf{r}^{\prime }\right) &=&\frac{1}{F}\sin
\delta _{0}\frac{z^{2}}{rr^{\prime }}\left[ j_{1}\left( k_{\mathrm{F}%
}r^{\prime }\right) \left( \gamma \left( \mathbf{r}\right) \cos \frac{\pi
\Phi }{\Phi _{0}}+\tilde{\gamma}\left( \mathbf{r}\right) \sin \frac{\pi \Phi 
}{\Phi _{0}}\right) \right. +  \label{Fr1r2_magn} \\
&&+j_{1}\left( k_{\mathrm{F}}r\right) \left( \gamma \left( \mathbf{r}%
^{\prime }\right) \cos \frac{\pi \Phi }{\Phi _{0}}+\tilde{\gamma}\left( 
\mathbf{r}^{\prime }\right) \sin \frac{\pi \Phi }{\Phi _{0}}\right) +  \notag
\\
&&+\sin \delta _{0}\left( j_{0}\left( k_{\mathrm{F}}d\right) -j_{0}\left( k_{%
\mathrm{F}}\sqrt{4z^{2}+d^{2}}\right) \right) \times  \notag \\
&&\left. \left( j_{1}\left( k_{\mathrm{F}}r\right) j_{1}\left( k_{\mathrm{F}%
}r^{\prime }\right) -y_{1}\left( k_{\mathrm{F}}r\right) y_{1}\left( k_{%
\mathrm{F}}r^{\prime }\right) \right) \right] ,  \notag
\end{eqnarray}

\begin{equation}
\tilde{\gamma}\left( \mathbf{r}\right) =\cos \delta _{0}j_{1}\left( k_{%
\mathrm{F}}r\right) +\sin \delta _{0}\left\{ y_{1}\left( k_{\mathrm{F}%
}r\right) \left( j_{0}\left( 2k_{\mathrm{F}}z\right) -1\right) -y_{0}\left(
2k_{\mathrm{F}}z\right) j_{1}\left( k_{\mathrm{F}}r\right) \right\} ,
\end{equation}%
where $\gamma \left( \mathbf{r}\right) $ is defined by (\ref{Gamma}), $\Phi
_{0}=\pi c\hbar /e$ is the flux quantum and $\Phi =\mathbf{HS}$ is the
magnetic flux through the triangle formed by vectors $\mathbf{r}_{0},$ $%
\mathbf{r}_{0}^{\prime }$, and the vector $\mathbf{d}$ connecting the
contacts. At $H=0$ the expression (\ref{FullCond_magn}) reduces to the
formula obtained earlier (\ref{FullCond}).

At $r_{0},r_{0}^{\prime }\gg 1/k_{\mathrm{F}}$, and $\delta _{0}\ll 1,$ Eq.(%
\ref{Gosc_magn}) takes the form%
\begin{gather}
\frac{\Delta G_{osc}\left( r_{0},d,H\right) }{G_{0}}\simeq -\frac{12\delta
_{0}z_{0}^{2}f\left( k_{\mathrm{F}}d\right) }{k_{\mathrm{F}}^{2}\left(
r_{0}r_{0}^{\prime }\right) ^{2}}\times  \label{osc} \\
\left[ \cos k_{\mathrm{F}}r_{0}\sin \left( k_{\mathrm{F}}r_{0}^{\prime }-%
\frac{\pi \Phi }{\Phi _{0}}\right) +\cos k_{\mathrm{F}}r_{0}^{\prime }\sin
\left( k_{\mathrm{F}}r_{0}-\frac{\pi \Phi }{\Phi _{0}}\right) \right] . 
\notag
\end{gather}

Similar oscillations in the electron local density of states have been
predicted in Ref.~\cite{Cano} for a system of two adatoms and an STM tip in
a plane perpendicular to a surface magnetic field.

If the period of the oscillations is known, the depth $z_{0}$ can be
determined using the following procedure: In the most convenient geometry of
the experiment the contacts should be placed so that the vectors $\mathbf{r}%
_{0},$ $\mathbf{r}_{0}^{\prime }$ and the normal to the sample surface are
situated in the same plane, i.e. the vectors $\mathbf{H}$ and $\mathbf{S}$
are parallel. For our illustration in Fig.\ref{model} that means the
coordinate $\mathbf{\rho }_{0}$ of the defect in the plane $xy$ is on the
line connecting the tips. In this case the relation between the period of
oscillations $\Delta H$ and the depth $z_{0}$ is \ very simple%
\begin{equation}
z_{0}=\frac{4\Phi _{0}}{d\Delta H}  \label{z0}
\end{equation}

Note that observation of the conductance oscillations (\ref{osc}) requires a
sufficiently strong magnetic field. Currently in low temperature STM the
magnetic field up to $15T$ is reachable \cite{Shvarts}. For example, in
order to observe the quote of period $\Delta H$ for $z_{0}=d=$20 nm it is
necessary to apply the field $H=5T.$ For typical metals, for which $\lambda
_{\mathrm{F}}\sim 0.1$ nm, for the distance between the contacts and the
defect $r${}$_{0}\gtrsim 10$ nm the amplitude of conductance oscillations
become very small $G_{osc}\sim G_{0}\left( \lambda _{\mathrm{F}%
}/r{}_{0}\right) ^{2}$ $\sim \left( 10^{-4}\div 10^{-5}\right) G_{0}$.
Therefore more suitable objects for application of proposed magnetic method
of determination of the defect position below the surface are
semiconductors, semimetals (Bi, Sb and their ordered alloys) where the Fermi
wave length $\lambda _{\mathrm{F}}\sim 10$ nm. Also the large amplitude $%
G_{osc}\sim \left( 10^{-2}\div 10^{-3}\right) G_{0}$ could be expected in
the metals of the first group, a Fermi surface of which has small cavities
with effective mass $m^{\ast }\simeq 10^{-2}\div 10^{-3}m_{0}$ ($m_{0}$ is
the mass of a free electron). As well a low temperature STM should be used
to avoid electron-phonon scattering on the electron trajectory.

Thus, in this paper we have investigated theoretically the conductance of
the system consisting of two close tunnel point contacts in the vicinity of
which the point defect is situated. In approximation of $s-$wave scattering
which is valid for short range scattering potential the oscillatory
dependence of conductance on the separation between the contacts and their
distances from the defect is studied. We proposed an alternative way that
allows to determine the depth of the subsurface impurity by measuring the
phase change in the conductance oscillation, arising when we change the
distance between the contacts of double-tip STM. Also it was obtained that
when the case of low magnetic field which is parallel to the surface of the
sample the depth of the subsurface impurity can be easily found from the
period of Aharonov-Bohm type oscillations of conductance, which arise in
this case.

\end{document}